\documentclass[12pt,preprint]{aastex}
\usepackage[twocolumn]{emulateapj5}
\usepackage{lscape}

\shorttitle{Ultraluminous X-ray Sources: NGC 1313 X-2}
\shortauthors{Zampieri et al.}

\begin{document}

\title{The Ultraluminous X-ray Source
       NGC 1313 X-2 (MS 0317.7-6647) and its Environment}

\author{Luca Zampieri, Paola Mucciarelli\altaffilmark{1},
    Renato Falomo}
\affil{INAF-Osservatorio Astronomico di Padova,
    Vicolo dell'Osservatorio 5, I-35122 Padova, Italy}

\author{Philip Kaaret}
\affil{Harvard-Smithsonian Center for Astrophysics, 60 Garden Street,
Cambridge, MA 02138}

\author{Rosanne Di Stefano}
\affil{Harvard-Smithsonian Center for Astrophysics, 60 Garden Street,
Cambridge, MA 02138}
\affil{Department of Physics and Astronomy, Tufts University, Robinson Hall,
Medford, MA 02155}

\author{Roberto Turolla}
\affil{Dipartimento di Fisica, Universit\`a di Padova, Via Marzolo 8, I-35131
Padova, Italy}


\author{Matteo Chieregato, Aldo Treves}
\affil{Dipartimento di Scienze, Universit\`a dell'Insubria, Via Valleggio 11,
I-22100 Como, Italy}

\altaffiltext{1}{Also at SISSA, Via Beirut 2-4, I-34014 Trieste, Italy}

\begin{abstract}
We present new optical and {\it Chandra} observations of the field
containing the ultraluminous X-ray source NGC 1313 X-2. On an ESO 3.6 m
image, the {\it Chandra} error box embraces a $R=21.6$ 
point-like object and excludes a previously proposed optical
counterpart. The resulting X-ray/optical flux ratio of NGC 1313 X-2 is
$\sim 500$. The value of $f_X/f_{opt}$, the X-ray variability history
and the spectral distribution derived from a re-analysis of the {\it
ROSAT\/}, {\it ASCA\/} and {\it XMM} data indicate a luminous X-ray
binary in NGC 1313 as a likely explanation for NGC 1313 X-2. If the
X-ray soft component observed in the {\it XMM} EPIC spectrum
originates from an accretion disk, the inferred mass of the compact
remnant is $\approx 100 M_\odot$, making it an intermediate mass black
hole. The derived optical luminosity ($L\approx 10^5 L_\odot$) is
consistent with that of a $\approx 15-20 M_\odot$ companion. The
properties of the environment of NGC 1313 X-2 are briefly discussed.
\end{abstract}


\section{Introduction}

First revealed by the {\it Einstein\/} Observatory (see
e.g. \citealt{fab89}), point-like, off-nuclear X-ray sources with
luminosities significantly exceeding the Eddington limit for one solar
mass are being progressively discovered in the field of many nearby
galaxies. To date, hundreds of such sources have been found in dozens of
galaxies, both ellipticals and spirals (e.g. \citealt{col02}). These
powerful objects, commonly referred to as ultraluminous X-ray sources
(ULXs), do not appear to have an obvious Galactic counterpart. Despite
some of them have been identified with supernovae or
background active galactic nuclei, the nature of most of these sources
remains unclear. X-ray spectra have been obtained with the {\it
Einstein\/}, {\it ROSAT\/}, {\it ASCA\/} and recently {\it
XMM-Newton\/} and {\it Chandra\/} satellites. Although the statistics
is rather poor, in many cases fitting with simple models indicates
that the spectral properties are consistent with those of Galactic
black hole binaries (e.g. \citealt{fosc02a}). About half of them show
some degree of variability in the X-ray flux. Among the various 
possibilities, the most favored
explanation is that ULXs are powered by accretion and that they are
somewhat special X-ray binaries, either containing an intermediate
mass black hole (BH) with $M_{BH}\ga 100 \, M_\odot$ (e.g. 
\citealt{colbert99}; \citealt{kaaret01}) or having beamed
emission toward us (e.g. \citealt{king01}; \citealt{kaaret03}). 
For a recent review on the properties of ULXs we refer to 
\cite{fabbiano03}.

Optical observations are of fundamental importance to better assess
the nature of these sources but they are still rather scarce (see
e.g. \citealt{cagn03}; \citealt{fosc02b}). Some ULXs have optical
counterparts in the Digitized Sky Survey or Hubble Space Telescope
images (e.g. NGC 5204 X-1; \citealt{rob01,goad02}) and some appear to
be embedded in emission nebulae a few hundred parsecs in diameter
\citep{pak02}.

NGC 1313 X-2 was one of the first sources of this type to be found. It
was serendipitously discovered in an {\it Einstein\/} IPC pointing
toward the nearby SBc galaxy NGC 1313 \citep{fab87}. Originally
included in the {\it Einstein\/} Extended Medium Sensitivity Survey as
MS 0317.7-6647, it is located $\sim 6'$ south of the nucleus of NGC
1313. \cite{sto95} investigated the nature of MS 0317.7-6647 on the
basis of X-ray, optical and radio observations. They identified a
possible optical counterpart and concluded that the source could be
either a Galactic isolated neutron star or a binary containing a
massive BH in NGC 1313. Spectral fits to {\it ROSAT\/} PSPC data
\citep{sto95,col95,mil98} yielded results consistent with many single
component models. {\it ASCA\/} observations \citep{pet94,mak00} are
described successfully by a multi-color disk blackbody (MCD) model,
representing thermal emission from a standard
accretion disk around a BH. A very recent analysis of a {\it XMM\/}
EPIC-MOS observation of NGC 1313 \citep{mil02} indicates that two spectral
components, soft and hard, are required to fit the spectrum of NGC
1313 X-2 and the normalization of the soft component yields
a conspicuous mass of the black hole $M_{BH}\ga 830 M_\odot$.

We present new optical\footnote{Based on observations collected at the
European Southern Observatory, Chile, Program number 68.B-0083(A).} 
and {\it Chandra} observations of NGC 1313 X-2, with the aim to shed
further light on its enigmatic nature. In \S~2 we present {\it
Chandra} data and re-analyze all the available X-ray observations of
NGC 1313 X-2. In \S~3 optical observations of the field of this ULX
are reported. Finally, the implications of our results on the nature
of NGC 1313 X-2 are discussed in \S~4.

\section{X-ray Data}

NCG 1313 X-2 was first observed by {\it Einstein} with the IPC
instrument in 1980. It was then pointed several times by {\it ROSAT\/}
(PSPC and HRI) between 1991 and 1998 \citep{sto95,col95,mil98,sch00},
by {\it ASCA\/} (SIS and GIS) in 1993 and 1995 \citep{pet94,mak00} and
by {\it XMM\/} (EPIC) in 2000 \citep{mil02}. Most recent data are from a 
2002 {\it Chandra\/} (ACIS-S)
pointing, reported here for the first time. The {\it Chandra\/}
observation began on 13 Oct 2002 and had a duration of 19.9 ks.  The
primary goal of the observation was to study sources near the center
of the galaxy, but the aim-point was adjusted to also place NGC 1313
X-1, NGC 1313 X-2, and SN 1978K on the S3 chip of the ACIS-S.
In addition, we present here a complete analysis of the {\it XMM} 
observation, including the EPIC-PN data, which were not considered by 
\cite{mil02}.

\subsection{X-ray astrometry}

An accurate determination of the X-ray position of NCG 1313 X-2 can be
obtained from the 2002 {\it Chandra} pointing, using the {\it Chandra}
aspect solution. Chandra data were extracted from the S3 chip on the
ACIS-S and subjected to standard processing and event screening.  No
strong background flares were found, so the entire observation was
used.  Because the source is $5\arcmin$ off axis, the point spread
function was fitted with an ellipsoidal Gaussian (1.9$''$ and 1.1$''$ 
along the two axes, rms values).  Also, the
pixel with the highest number of counts is offset by 0.8$''$ from the
center of the fitted ellipse. Taking these uncertainties into account,
we conservatively estimate a positional error of 0.7$''$ (1-$\sigma$).
The final {\it Chandra} position is: $\alpha=$ 03h 18m
22.27s$\pm$0.12s, $\delta=$ -66$^0$ 36$'$ 03.8$'' \pm$0.7$''$.

In order to check the accuracy of the {\it Chandra} aspect solution,
we exploited the presence in the field of view of a quite peculiar
supernova, SN 1978K, that shows powerful radio and X-ray emission. The
{\it Chandra} position of SN 1978K is $\alpha=$ 03h 17m 38.69s,
$\delta=$ -66$^0$ 33$'$ 03.6$''$ (J2000), within $0.46''$ from the
accurate (0.1$''$) radio position of \cite{ryd93}. This is consistent
with the expected {\it Chandra} aspect accuracy.

The position of NGC 1313 X-2 was previously determined from the {\it
ROSAT\/} HRI (\citealt{sto95,sch00}) and {\it XMM} EPIC-MOS
\citep{mil02} images. Typical 1-$\sigma$ error boxes 
are $\sim 3''$ for {\it ROSAT\/} HRI and $\sim 2''$ for
{\it XMM} EPIC-MOS. The {\it ROSAT\/} and {\it XMM} positions and
corresponding error boxes are summarized in Table \ref{tab0}.

\subsection{X-ray spectrum and lightcurve}

We analyzed the {\it XMM} EPIC data from both the MOS and PN cameras
(operated with the medium filter). The EPIC-PN spectrum is reported
here for the first time and was extracted directly from the
observation data file because the automatic pipe-line processing
failed to produce a standard event list for the EPIC-PN camera. Both
reduction procedures ({\tt epchain} and {\tt epproc}) were used to
extract the data obtaining similar results (differing typically by a
few percents). Data screening, region selection and event extraction
were performed using standard software (XMMSELECT v 2.43.2). An
analysis of the MOS and PN light curves shows that solar flares are
present in both datasets. They were filtered out using the standard
criterion (total off-source count rate above 10 keV $<$ 5 counts
s$^{-1}$ for MOS and $<$ 15 counts s$^{-1}$ for PN). We extracted the
source counts from a circle of 40$''$ and 30$''$ for the MOS and PN
cameras, respectively. The proximity of the source to one of the CCD
edges in the EPIC-MOS data requires some care. We eliminated the area
of a box aligned and superimposed to the CCD boundary to avoid
contamination from bad pixels close to the source. The background was
selected from a circle of 60$''$ in a nearby source-free region of the
same CCD. Ancillary and response files were produced using the
appropriate XMMSELECT tasks. Data were grouped to require at least 20
counts per bin for the MOS data and 40 counts per bin for the PN data,
and were then analyzed and compared with different models using XSPEC
(v 11.2.0). To minimize the effects of possible relative calibration
uncertainties, the fit of the MOS1, MOS2 and PN spectra were performed
with an overall normalization constant (those of the two MOS cameras
differ by $\sim 10$\%, while that of the PN instrument is larger by
$\sim 25$\%). The count rates are 0.08 counts s$^{-1}$ for the MOS
cameras and 0.25 counts s$^{-1}$ for the PN.

In order to reconstruct the X-ray variability history of NGC 1313 X-2,
we have carefully re-analyzed also the {\it ROSAT\/} and {\it ASCA\/}
observations. Extraction regions for the {\it ASCA} SIS data were
chosen with care to avoid contamination from the CCD edges and SN
1978K. Spectra were grouped to require at least 15 counts per bin for
the {\it ROSAT} data and 20 counts per bin for the {\it ASCA} data.

The results of the spectral analysis are listed in Table \ref{tab1}.
The statistics of the {\it XMM} EPIC spectrum is significantly
improved including the EPIC-PN data (the PN camera has almost twice
more counts than each single MOS instrument). An absorbed power-law
does not provide a satisfactory fit of the joint MOS1, MOS2 and PN
spectra ($\chi_{red}^2=1.2$ for 249 {\it d.o.f.}). Two components
models provide a significant improvement over single component
ones. The best fit is obtained with an absorbed soft, thermal
component plus a power-law. Adding a MCD model to the power-law
results in an improvement of the fit which is significant at the $\sim
4.5/5$ ({\tt epchain}/{\tt epproc}) $\sigma$ level.  Figure
\ref{fig2b} shows the results for a MCD+power-law fit. The resulting
best fitting parameters are $kT = 200_{-40}^{+50}$ eV, $\Gamma =
2.23_{-0.09}^{+0.15}$ and $N_H = 3.13_{-0.37}^{+0.92}
\times 10^{21}$ cm$^{-2}$ for the inner disk temperature, photon index
and column density, respectively (see Table \ref{tab1}). There are
residuals in the fit (mainly in the EPIC-PN spectrum) that suggest the
possible presence of emission lines.
We emphasize that the EPIC-PN data provide marginal evidence for the
presence of a soft component even at low metallicities. Reducing the
abundances of the absorbing gas at 0.5 solar, a simple power-law fit
of the EPIC-PN data has $\chi_{red}^2=1.24$ (89 {\it d.o.f.}), while a
MCD+power-law fit gives $\chi_{red}^2=1.15$ (87 {\it d.o.f.}).
Finally, it is worth noting that the value of $kT$ is $\sim 25$\%
larger (although consistent within 1-$\sigma$) than that derived by
\cite{mil02}.

\begin{figure}
\epsscale{0.6}
\plotone{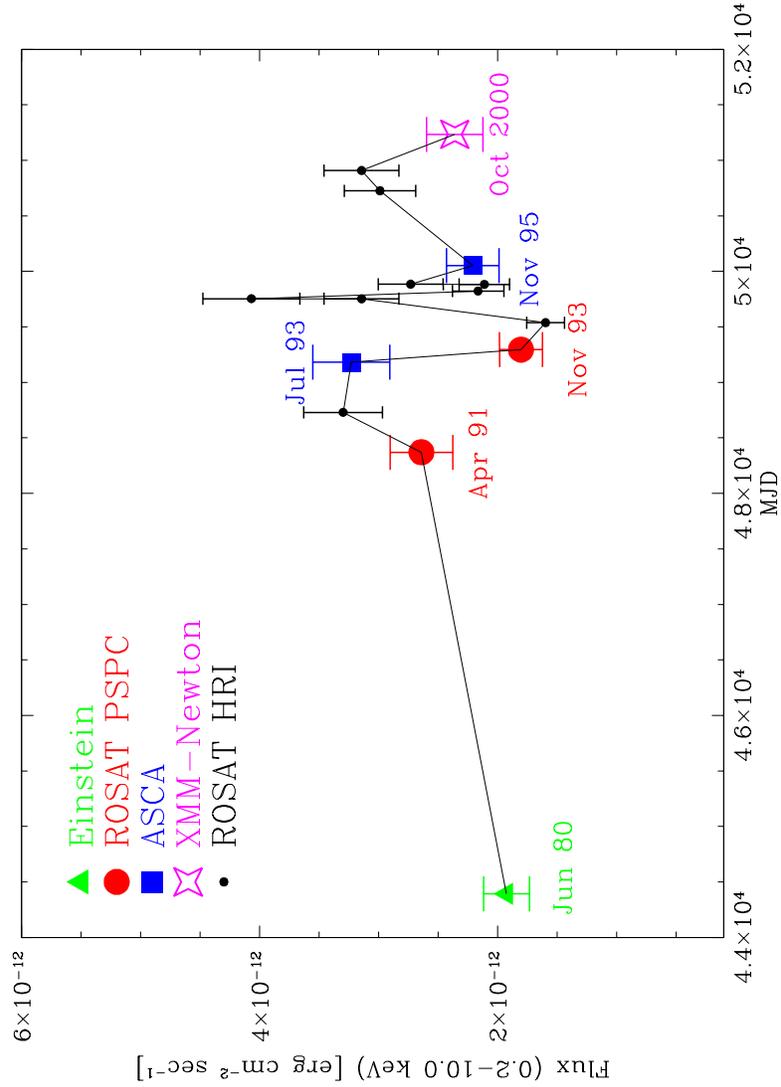}
\caption{The 0.2-10 keV light curve of NGC 1313 X-2 from
all presently available data. Fluxes were de-absorbed and, when
necessary, extrapolated in the 0.2--10 keV interval using the web
interface to PIMMS (see Table \ref{tab1}). For the {\it Einstein\/}
IPC observation, we adopt the value reported by
\cite{gioia90}. For the {\it ROSAT\/} PSPC and {\it ASCA\/} GIS+SIS
observations, the fluxes were derived from the best fit parameters of
the power-law model reported in Table
\ref{tab1}. For the {\it ROSAT\/} HRI data \citep{sch00} a power-law
spectrum with $\Gamma=2$ and $N_H=3 \times 10^{21}\, {\rm cm}^{-2}$
was assumed (in agreement with the best fit parameters derived from
the spectral analysis). For the {\it XMM} EPIC data, the flux is
calculated from the best fit MCD+power-law model reported in Table
\ref{tab1}.
\label{fig1}}
\end{figure}

\begin{figure}
\epsscale{0.8}
\plotone{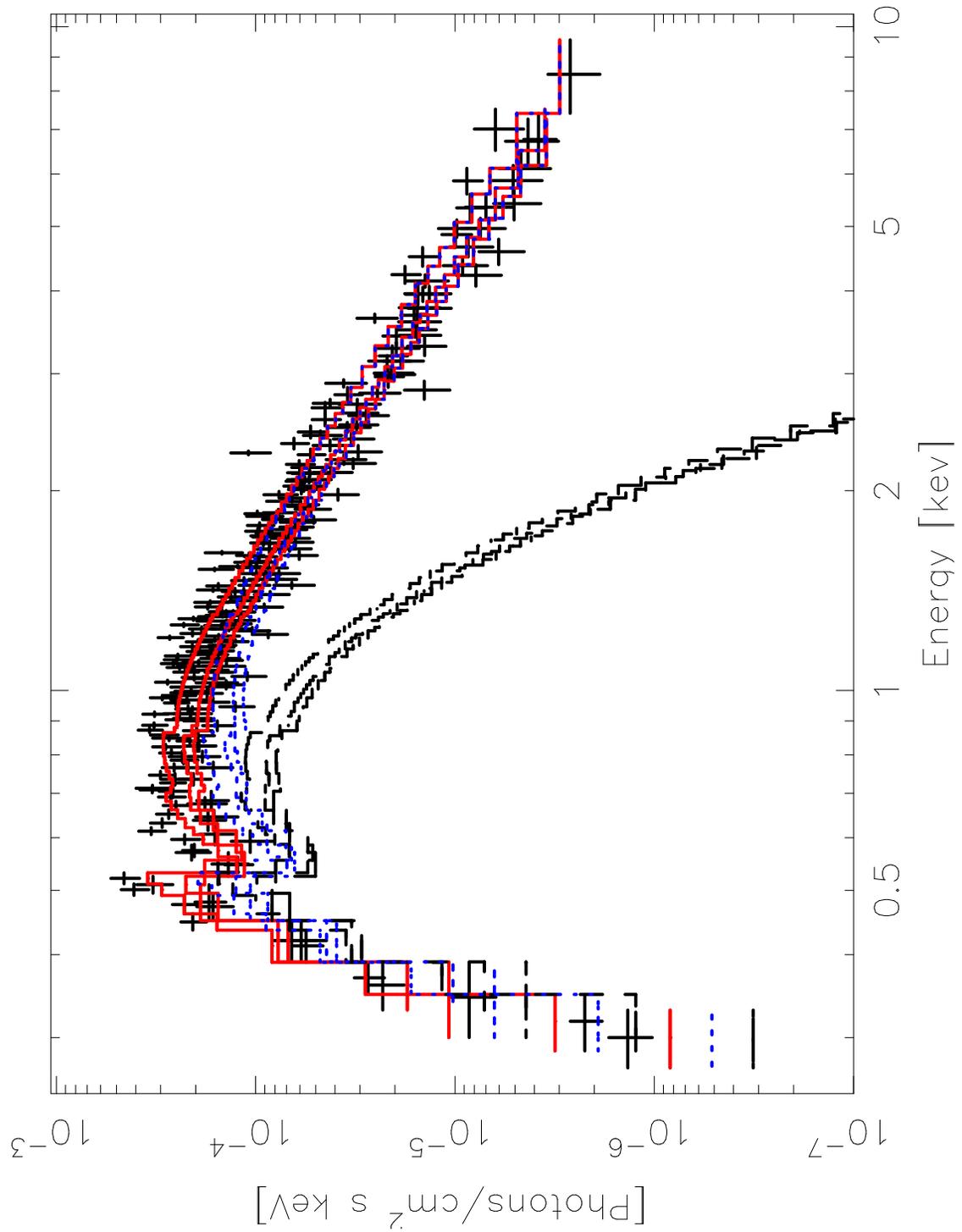}
\caption{X-ray spectrum of NGC 1313 X-2 from the {\it XMM} EPIC-MOS and PN
cameras. The solid line represents the combined best fitting model
spectrum, while the dashed and dotted lines are the MCD and power-law
components respectively.\label{fig2b}}
\end{figure}

In Figure \ref{fig1} we report the X-ray flux derived from all the
available observations of NGC1313 X-2. The fluxes were consistently
derived from the best fit parameters of the X-ray spectral analysis
reported in Table \ref{tab1}. An approximate estimate of the errors,
based on counting statistics, is 5-10\%. The {\it Chandra} point is
not included in Figure \ref{fig1} because of pile-up problems. The
unabsorbed 0.2--10 keV flux from both the {\it XMM} EPIC-MOS and PN
instruments agree within $\sim 20\%$ and give an average value of
$2.4 \times 10^{-12}$ erg cm$^{-2}$s$^{-1}$. This value is lower 
by a factor $\sim 2$ than that estimated
by \cite{mil02}. Variability of up to a factor 2 on a timescale of
months is clearly present and it is reminiscent of the behavior
observed in Galactic X-ray binaries. If uncertainties in the
best-fitting spectral parameters are taken into account, the amplitude
of variability is reduced but not eliminated. This suggests that a
compact object is present in NGC 1313 X-2. If the emission is isotropic
and the distance is that of the host galaxy ($\simeq 3.7$ Mpc;
\citealt{tul88}), the X-ray luminosity in the 0.2--10 keV range is
$L_X \simeq (3 - 6 \pm 0.5) \times 10^{39}$ erg s$^{-1}$. If at
maximum the source radiates at the Eddington limit $L_{Edd}$, the BH
mass is $\sim 50 \, M_\odot$. Sub-Eddington accretion would imply an
even larger mass.

Interestingly, comparing data from the first and second epoch {\it
ASCA} observations, the X-ray flux of NGC 1313 X-2 appears to increase
with increasing spectral hardness (see Table \ref{tab1}). This
behavior is similar to that observed in the ULXs of the Antennae
galaxy \citep{zez02} and is opposite to what is usually seen in
Galactic BH X-ray binaries (e.g. Cyg X-1).

\section{Optical Observations}

Optical images of the field of NGC 1313 X-2 in the $R$-band (Bessel
filter) were taken on 16 January 2002 with the 3.6 m telescope of the
European Southern Observatory (ESO) at La Silla (Chile). We used
EFOSC2 with a Loral/Lesser CCD of 2048$\times$2048 pixels yielding a
field of view of $\sim 5' \times 5'$ at a resolution of
0.314$''$/pixel (re-binned by a factor 2). The night was clear with a
seeing of about 1$''$.  Four images were obtained for a total exposure
time of 1320 s. Standard reduction of the data (including bias
subtraction and flat-field correction) was performed within the IRAF
(v 2.12) environment.

A spectrum of one of the field objects (object A; see below) was
secured on the same night. We performed low-resolution (13.4 A,
grism\#4) spectroscopy for a total exposure time of 1200 s. After
applying standard corrections and sky subtraction, cosmic rays were
removed and the spectrum was corrected for atmospheric extinction. At
the time of the optical observations object A had already been imaged
at the 1.1 m Las Campanas telescope by \cite{sto95} and was considered
a possible counterpart of NGC 1313 X-2. Although the new accurate {\it
Chandra} position rules out an association with this object (see
below), the spectrum can be used to gain insight on the properties of
a surrounding nebula, possibly associated with the X-ray source (see
e.g. \citealt{pak02}).

\subsection{Astrometry and photometry of field objects}

Our four ESO images were astrometrically calibrated using an IRAF task
({\sc PLTSOL}) and performing a polynomial interpolation starting from
the positions of GSC2 ESO field stars. The internal accuracy of this
procedure was estimated comparing the actual positions of a number of
GSC2 stars not used for astrometric calibration with the positions
contained in the catalog. The accuracy is 0.3$"$ (1-$\sigma$). The four
calibrated images were then summed together and the resulting image is
shown in Figure \ref{fig3}.

In order to check for the relative systematics between the optical and
X-ray astrometric calibrations, we used again the position of SN 1978K. 
This supernova is inside the {\it Chandra} field of view but outside our
optical image. Thus, we analyzed also an archival image of SN 1978K
(from the Padova-Asiago Supernova Archive) taken on 13 September 1999
with the same telescope and a similar instrumental set-up (ESO
3.6m+EFOSC/2.9+R\#642, exposure time 180 s). After calibrating the
archival image, the position of SN 1978K is $\alpha=$ 03h 17m 38.605s,
$\delta=$ -66$^0$ 33$'$ 03.13$''$ (J2000). This is within 0.28$''$
from the radio position of \cite{ryd93}, improving significantly upon
the previous optical position by the same authors. The difference
between the centroids of the optical and {\it Chandra} positions of SN
1978K is 0.69$''$ ($\alpha_{opt}-\alpha_X=$ -0.085s,
$\delta_{opt}-\delta_X=$ -0.47$''$). Although this difference is small
and comparable with the statistical errors, we decided to apply this
correction to the {\it Chandra} position of NGC 1313 X-2 to eliminate
any systematic error between the optical and X-ray astrometric
calibrations. The resulting {\it Chandra} position of NGC 1313 X-2 is
reported in Table \ref{tab0}.

The photometry of the objects in our optical image was performed
calibrating the frame with the $R$-band magnitudes of 23 stars from
the SuperCosmos Sky Survey \citep{humbly01} homogeneously distributed
over the field of view. The internal accuracy of this calibration is
0.2 mag.  Aperture (5$''$ radius) magnitudes are reported in Table
\ref{tab0}.

\subsection{Spectroscopy of the emission nebula}

The two-dimensional spectrum of the field around object A (Figure
\ref{fig6}) shows clear emission lines extending for tens of arcsecs
from east to west, confirming the existence of an extended ($\sim$ 400 pc)
optical emission nebula  that was first found in deep
H$_\alpha$ images by \cite{pak02}. A one-dimensional spectrum of the
field was extracted over an aperture of 0.9$''$ (3 pixels) from two
regions east-ward and west-ward of the position of object A and
adjacent to it (Figure \ref{fig4}). Wavelength and relative flux
calibration were applied to the data. The one-dimensional spectrum
shows strong emission lines of H$_\alpha$, H$_\beta$, [SII]
$\lambda\lambda$ 6717--6731 A, [OI] $\lambda$ 6300 A and [OIII]
$\lambda\lambda$ 4959-5007 A. The shift of the centroid of the lines
($\sim 10$ A) is consistent with the recession velocity of the galaxy
and indicates that the emission nebula is located in NGC 1313.

It is worth emphasizing the abrupt change in the absolute and relative
intensity of the emission lines from east to west, indicating
variations in the physical conditions and/or geometry of the emission
nebula. In particular, strong emission from [OIII] is present on the
east side but almost absent on the west side, while emission from
H$_\alpha$, [SII], [OI] and other elements is present on the west side
but weaker or absent on the east side. A careful inspection of the
line intensity profiles (in particular those of H$_{\alpha}$ and
[SII]) reveals a fairly symmetric, broadly peaked profile, centered at
$\sim 2''$ west of the position of object A, and a weaker, roughly
constant intensity component extending in the east direction.

\section{Discussion}

\begin{figure}
\epsscale{0.8}
\plotone{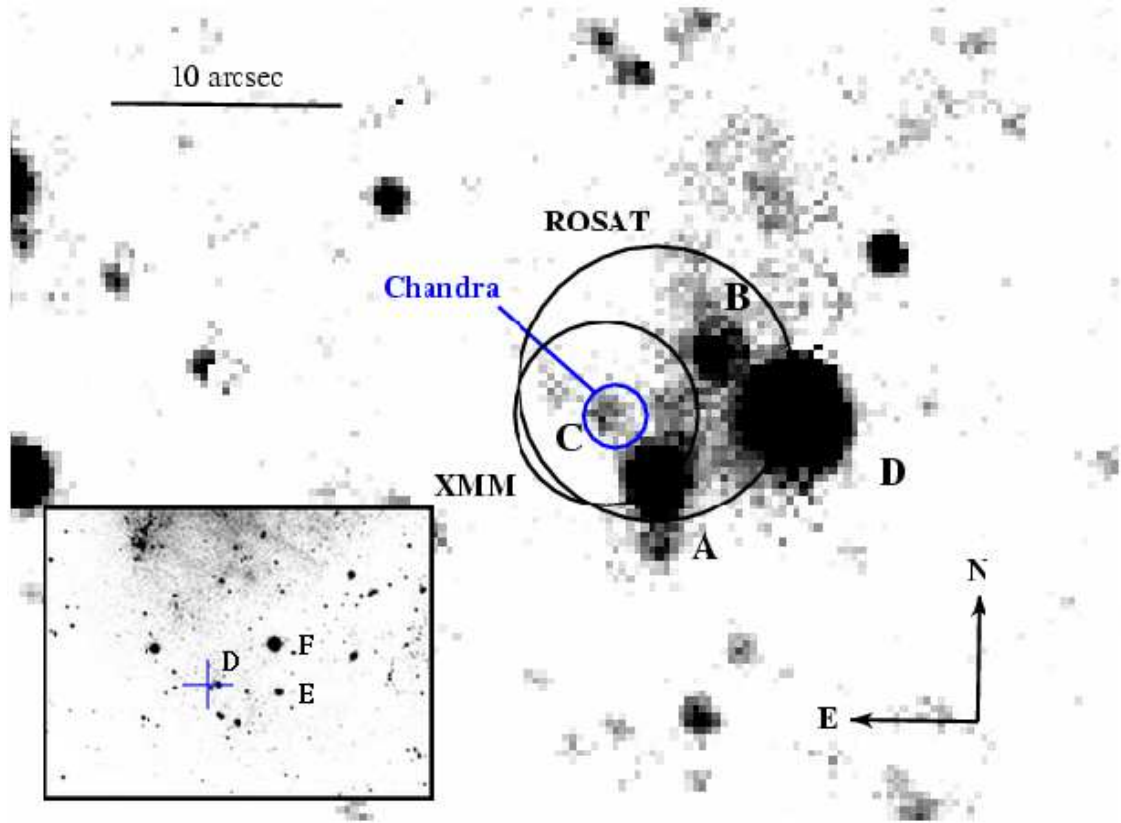}
\caption{ESO 3.6m $R$-band (Bessel filter) image of the field of NGC 1313 X-2.
The circles show the {\it ROSAT\/} HRI, {\it XMM\/} EPIC-MOS and {\it
Chandra} ACIS-S positions. The estimated 90\% confidence radii are
6$''$ for HRI, 4$''$ for EPIC-MOS and 1.4$''$ for ACIS-S. Labels A, B,
C and D mark the four field objects inside or close to the X-ray error
boxes. The insert at the bottom-left shows a larger portion of the
image with the position of the X-ray source ({\it
cross}). \label{fig3}}
\end{figure}

Our {\it Chandra} position of NGC 1313 X-2 (Table \ref{tab0}) is shown in
Figure \ref{fig3}, together with the {\it ROSAT\/} HRI \citep{sch00}
and {\it XMM} EPIC-MOS \citep{mil02} error boxes, overlaid on our ESO
image. All measurements are consistent within 1-$\sigma$. The distance
of the centroids of objects A, B and D with respect to the
{\it Chandra} position is 3.6$''$, 4.1$''$ and 7.3$''$,
respectively. Even taking into account the statistical error on the
optical positions (0.3$''$), these three objects can be ruled out at a
significance level of at least 3-$\sigma$. On the other hand, object C
is inside the Chandra error box and its position coincides within
1-$\sigma$ with that of NGC 1313 X-2, making it a likely counterpart.

From the maximum absorbed X-ray flux of NGC 1313 X-2 ($f_X\sim 2\times
10^{-12}$ erg cm$^{-2}$ s$^{-1}$) and optical magnitude of object C
($R=21.6$), we estimate $f_X/f_{opt}\sim 500$. This value is very
high, in agreement with the suggestion by \cite{cagn02} that
ULXs can be selected on the basis of their large $f_X/f_{opt}$. Only
Isolated Neutron Stars (INSs), heavily obscured AGNs and luminous
X-ray binaries can reach such large values of the X-ray/optical flux
ratio. INSs are extreme in this respect, with $B \approx 25$ optical
counterparts and typical X-ray-to-optical flux ratios $\ga 10^5$ (see
e.g. \citealt{kapl03})
The presence of the relatively bright ($R\sim 21.6$) object C in the {\it
Chandra} error box makes this possibility unlikely. Furthermore, known
INSs exhibit different spectral properties with no significant
variability. On the other hand, a heavily obscured AGN is expected to
have a rather hard X-ray spectrum and to emit significantly in the
near-infrared (see e.g. \citealt{brusa02}). Given the X-ray luminosity
of  NGC 1313 X-2, an infrared magnitude $K\approx 12$ is expected 
were it an obscured AGN. The lack of any
IR counterpart on a K image of the 2MASS All Sky Image Service down to
a limiting magnitude $K\simeq 14$ (10-$\sigma$) and the softer X-ray
spectrum of NGC 1313 X-2 make this possibility unlikely, although a
low resolution optical spectrum of object C is definitely required to 
settle this issue. Our accurate {\it Chandra} position and
optical identification favor a very luminous X-ray
binary in NGC 1313 as the likely explanation for NGC 1313 X-2. As a
reference, the X-ray/optical flux ratio of persistent BH binaries at
maximum is $\gtrsim 10-100$, while that of soft X-ray transients in
outburst can reach $2000$ (\citealt{mas97}). This is in line with the
alleged binary nature of ULXs and is consistent with the observed properties
of this source, such as the X-ray variability and the observed X-ray
spectrum, including the presence of a soft component probably produced
by an accretion disk.

If indeed NGC 1313 X-2 is a black hole binary, the X-ray spectral
parameters, in particular the temperature of the MCD fit (hereafter
referred to as $T_{MCD}$), can be used to estimate the BH mass. 
This is similar to what done by \cite{mil02} using the normalization of the
MCD fit and, as discussed below, we reach similar conclusions. 
The effective temperature of a standard accretion disk depends on radius
as $T^4=(3GM_{BH}{\dot M}/8\pi\sigma r_{in}^3)(r_{in}/r)^{3/4}
[1-(r_{in}/r)^{1/2}]$, where ${\dot M}$ is the accretion rate and
$r_{in}$ is the innermost disk radius
(e.g. \citealt{frakira}). Assuming that $T_{MCD}$ represents an
estimate of the maximum disk temperature, it is
$(3GM_{BH}{\dot M}/8\pi\sigma r_{in}^3)^{1/4}=\alpha T_{MCD}$, with
$\alpha\simeq 2$. Neglecting relativistic corrections and assuming
that the disk terminates at the innermost stable circular orbit of a
Schwarzschild BH, it is: $M_{BH}/M_\odot=({\dot M}c^2/L_{Edd}) f^4
(\alpha T_{MCD}/1.5\times 10^7 \, {\rm K})^{-4}$, where $f$ is a color
correction factor  ($f\sim 1.6-1.7$,
\citealt{shim95,zamp01}). Given the strong dependence of $M_{BH}$ on
temperature, any uncertainty in the accretion physics and radiative
transfer may induce significant errors in the resulting value of
$M_{BH}$. So, the inferred spectroscopic measurement of the BH mass
should be taken simply as an approximate estimate.
The low inner disk temperature obtained from the two-components fit to
the {\it XMM} EPIC spectrum ($kT \sim$ 200 eV) implies $M_{BH}
\approx 90 f^4 \, M_\odot$ ($\alpha\simeq 2$) for Eddington limited
accretion, somewhat larger than that derived from the flux.  This
result removes the need for a rapidly spinning BH (invoked by
\citealt{mak00} from an analysis of the {\it ASCA\/} data) and agrees
with the conclusion of \cite{mil02} that NGC 1313 X-2 contains an
intermediate mass BH. The large inferred BH mass does not require
beamed emission. Then, the estimated accretion rate (assuming 10\%
efficiency) is ${\dot M} \sim 10^{-7} \, M_\odot
\, {\rm yr}^{-1}$, forcing the mass reservoir to be a companion star.

From the apparent magnitude ($R=21.6$) and absorption ($A_R
\simeq 1.7$, computed from the X-ray best fitting column density
$N_H\sim 3\times 10^{21}$ cm$^{-2}$; \citealt{bol78}) of object C, we
estimate an absolute magnitude $M_R\simeq -7.9$ and a luminosity in
the range $\sim 7\times 10^4-10^6 L_\odot$, depending on the adopted
bolometric correction. If this originates from the companion star, the
inferred luminosity is consistent with a $\approx 20 M_\odot$ main
sequence star or a $\sim 15-20 M_\odot$ evolved OB supergiant
(e.g. \citealt{bd84}), making NGC 1313 X-2 a high-mass X-ray
binary. In luminous Galactic X-ray binaries, the reprocessed optical
emission from the disk may be significant. Assuming that 20--30\% of
the X-ray flux produced in the innermost part of the accretion disk
intercepts the outer regions, for realistic values of the albedo ($\ga
0.9$; e.g. \citealt{djvna96}) few percents of the X-ray luminosity
($\approx 10^{38}$ erg s$^{-1}$) can be absorbed and re-emitted in the
optical band. Characteristic emission lines of X-ray-ionized H, He or
N, typically seen in luminous Galactic X-ray binaries should then be
detectable in the optical spectrum. Also X-ray heating of the
companion star itself may contribute to the optical emission. If the
optical luminosity comes in part from X-ray re-irradiation, the mass
of the companion would be lower. Taking $M_2 \sim
20 M_\odot$ as an upper limit for the mass of the companion, the mass
ratio of the binary is $q=M_2/M_{BH}\la 0.4 f^{-4} \ll 1$.  Writing
the binary separation as $a=2.16 R_2 [q/(1+q)]^{-1/3}$ \citep{pac71},
the orbital period of the system is $P\ga 0.15 (R_2/100 \,
R_\odot)^{3/2} f^2 (M_{BH}/50 \, M_\odot)^{-1/2}$ yr.
According to \cite{king01}, the system should not be a persistent 
X-ray source.

Although difficult to reconcile with
the properties of the environment surrounding NGC 1313 X-2 (see
below), we can not rule out also that the optical emission detected in
the {\it Chandra} error box originates from a stellar cluster (see
e.g. the case of a ULX in NGC 4565; \citealt{wu02}), in which case NGC
1313 X-2 may be a low-mass X-ray binary in the cluster.

The mass accretion rate required to produce the observed luminosity
may in principle be provided by Roche-lobe overflow from an evolved
companion or by a wind from a supergiant. In the first case,
evolutionary swelling of the companion keeps pace with the increase in
Roche lobe size and the system remains self-sustained: accretion is
likely to proceed through a disk. In the second case, assuming 10\%
accretion efficiency and that the BH can capture $\sim 1\%$ of the
mass outflow, the wind must be very powerful (${\dot M} \sim 10^{-5}
\, M_\odot \, {\rm yr}^{-1}$). A lower efficiency would require too
high a gas supply, so a disk is needed even in a wind-fed system. In
this case, however, the disk is probably much smaller than in a
Roche-lobe overflow system and the optical emission dominated by the
supergiant. On the other hand, in a Roche-lobe overflow system, an
extended, possibly re-irradiated accretion disk should contribute
heavily in the UV and $B$ bands, producing strong emission
lines. Thus, the two modes of mass supply are likely to be
distinguishable by optical spectroscopy.

\begin{figure}
\epsscale{0.8}
\plottwo{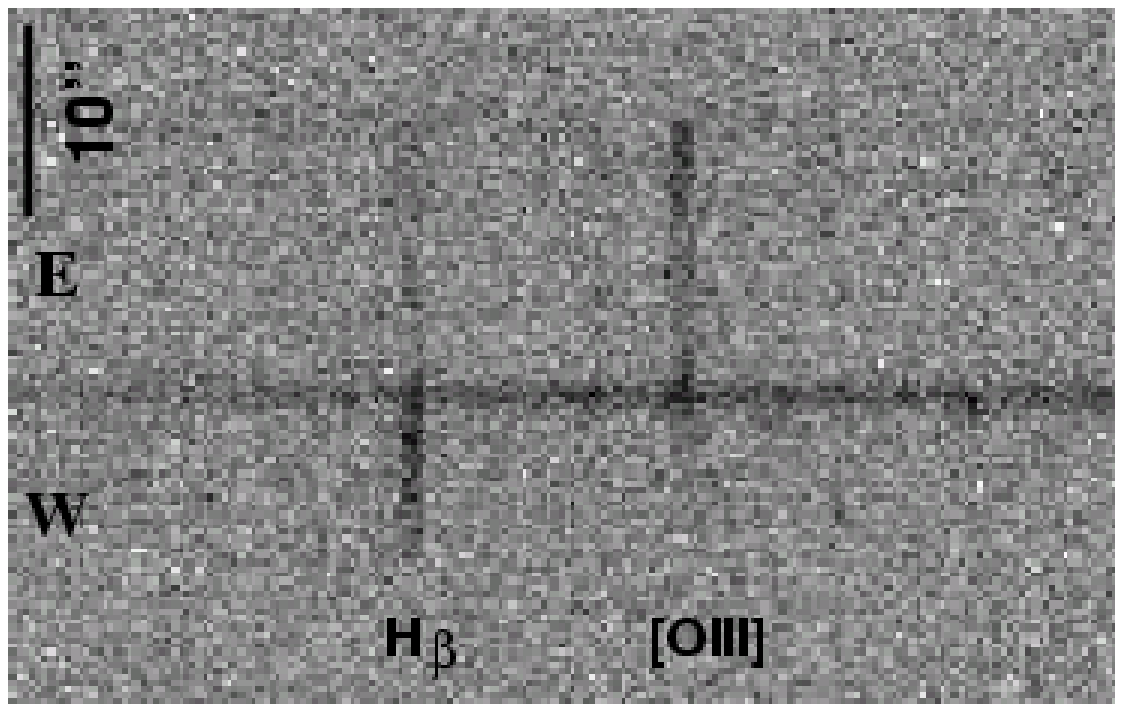}{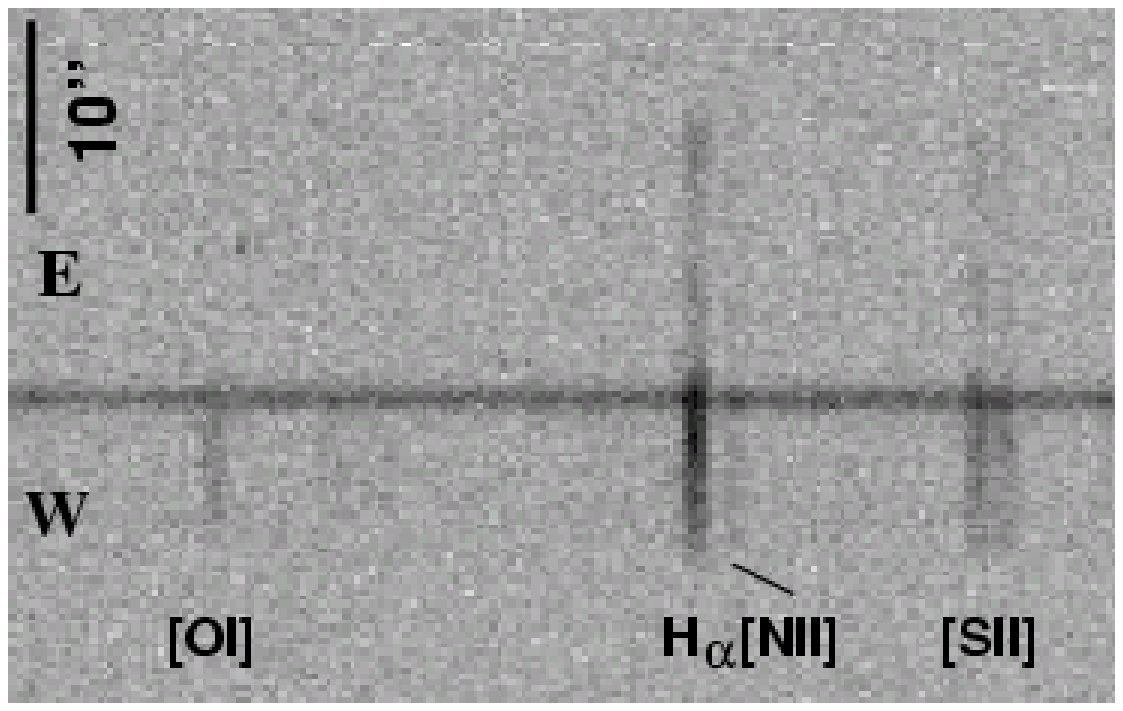}
\caption{Two-dimensional spectrum (ESO 3.6m+EFOSC2+grism\#4) of
the field around object A. The slit (1.2$''$) is oriented in the
east-west direction. The wavelength intervals are 4500--5300 A
(left panel) and 6150--6900 A (right panel). \label{fig6}}
\end{figure}

\begin{figure}
\epsscale{0.75}
\plotone{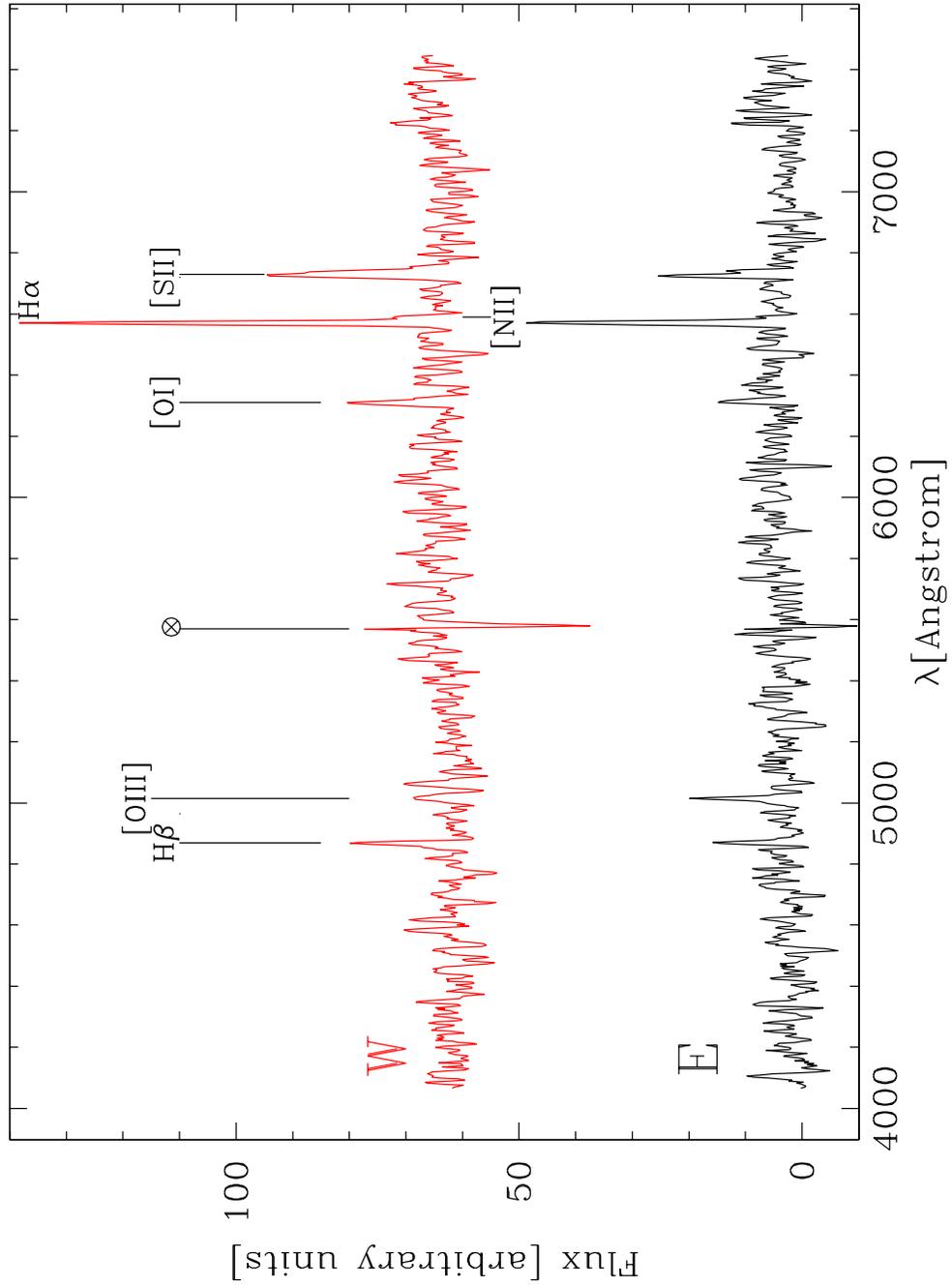}
\caption{One-dimensional spectrum ($F_\lambda$) of the nebula around
NGC 1313 X-2. The lower (upper) spectrum is extracted in a region
east-ward (west-ward) of the position of object A (see text for
details). The symbol $\otimes$ marks a residual contamination from an
emission line of the sky. \label{fig4}}
\end{figure}

\begin{figure}
\epsscale{0.8}
\plotone{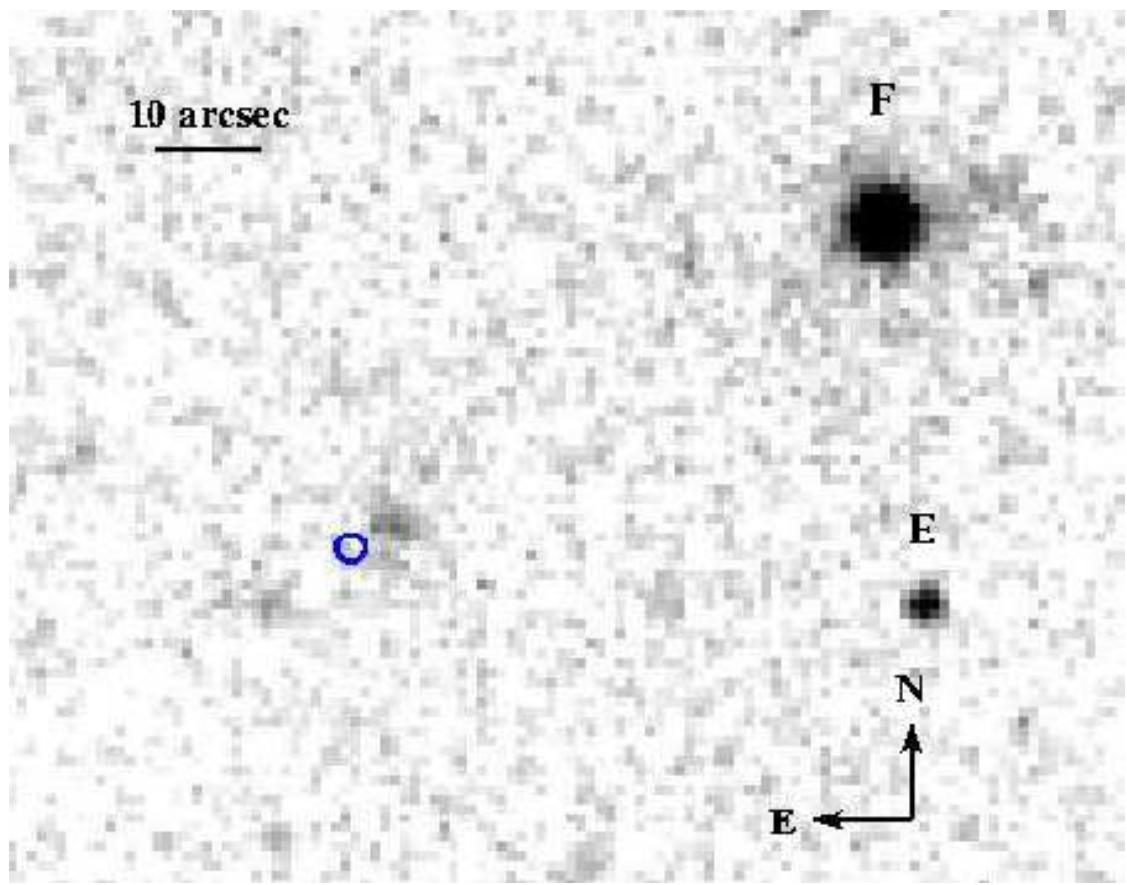}
\caption{UV-band
(UVW1 filter, 1800--3200 A) exposure of the field of NGC 1313
X-2, obtained with the {\it XMM\/} Optical Monitor. The circle shows
the {\it Chandra} position (90\% confidence level). A region of
possible diffuse emission is visible NW of the {\it Chandra\/} error
box.\label{fig7}}
\end{figure}

We now turn to discuss how our optical observations can be used to
constrain the environment of NGC 1313 X-2. Figures \ref{fig6} and
\ref{fig4} reveal that NGC 1313 X-2 is likely to be associated with an
optical emission nebula, recognizable also in a H$_\alpha$ image taken
by \cite{pak02}. From the velocity (80 km s$^{-1}$) and flux of
H$_\beta$, they derive an impressive mechanical energy of $3-10 \times
10^{52}$ erg for the expanding ionized gas, and suggest that the
nebula is inflated by a relativistic jet from NGC 1313 X-2. Our
measured ratio of [SII]/H$_\alpha$ ($\sim 0.5$) is consistent with
that expected from a shock-ionized supernova remnant, a stellar
wind-shocked nebula or diffuse ionized gas \citep{mat97}. However, the
inferred diameter and energy of the nebula are too large to be
consistent with a single supernova event, unless it was produced by a
hypernova similar to SN 1998bw (see e.g. \citealt{iwa98}). In fact, it
could be the result of several explosion events (multiple supernova
remnant) or be originated by the intense wind of hot stars, possibly
the parent stellar association of NGC 1313 X-2. As discussed in the
previous section, the nebula appears to have some internal structure:
a comparatively brighter, fairly symmetric component west of the
position of NGC 1313 X-2 and a weaker, slightly elongated one
extending in the east direction. The brighter part of the nebula has
[SII]/H$_\alpha$=0.58, the weaker one has [SII]/H$_\alpha$=0.44 and
intense [OIII] emission. Different possibilities may explain the
irregular appearance of the nebula. As suggested by \cite{pak02}, the
varying line intensity may be caused by reprocessed emission from the
X-ray ionized interstellar medium where the physical conditions (in
particular the density) vary on a scale $\sim 100$ pc.  However, the
nebular emission may also arise from two physically distict
components: a wind-shocked nebula produced by a possible parent
stellar association of NGC 1313 X-2 and a multiple supernova
remnant. This hypothesis seems to be confirmed also by the marginal
detection of (possibly extended) UV emission in an image of the {\it
XMM} Optical Monitor (see Figure
\ref{fig7}), in coincidence with the brighter component. Clearly, the
weaker component may still be a jet-inflated nebula, as suggested by
\cite{pak02}.
Finally, we note that, although NGC 1313 X-2 is somewhat hotter and
much more luminous, the [OIII] signature in the eastern portion of the
nebula is reminiscent of predictions for the radiation-limited nebulae
around supersoft sources (\citealt{distefano95}; \citealt{chiang96}).

It is interesting to note that object A, the possible association of
which 
with NGC 1313 X-2 was discussed by \cite{sto95}, lies very close both
to the {\it Chandra} error box and the point where the intensity of
the nebular emission lines suddenly changes. The continuum spectrum of
this object, obtained after subtracting off the emission line spectrum
of the nebula in the adjacent regions, was compared to template
stellar spectra \citep{pick98} and turns out to be in fair
quantitative agreement with that of a G-M supergiant, but not with
that of late-type dwarfs. This result is independent of reddening.
The absolute magnitude of object A is $M_R \simeq -9.7$, and the
luminosity $L \sim 5 \times 10^{5} L_{\odot}$ (assuming a bolometric
correction appropriate for a K star). Object A may be a very massive
($\sim 30 M_{\odot}$) G-M supergiant of radius $\sim 1000 R_{\odot}$
or a cluster in NGC 1313. The first interpretation would support the
conclusion that the region in which NGC 1313 X-2 is located is an
active star forming region, in which the initial mass function is
top-heavy. On the other hand, the second possibility appears more
likely because $\sim 30 M_{\odot}$ massive stars are extremely
rare. However, the red color would indicate a rather evolved cluster
that would be projected by chance on the active star forming
environment in which NGC 1313 X-2 appears to be embedded.


A crucial question is how a binary system containing an intermediate
mass BH may have formed (see e.g. \citealt{vdmarel03}). The BH
progenitor must have been rather massive. This is consistent with the
fact that NGC 1313 is likely to have lower than solar metallicity
($Z\sim 0.5$; \citealt{zkh94}) and hence mass loss was less
intense. Such a massive BH may have formed through direct collapse
without producing a supernova. In this way, if the system was born as
a binary, it may have survived after the collapse of the
primary. Although less likely, it is also possible that the companion
might have been captured from a nearby stellar association. In this
case, it is not possible to exclude that the BH may have formed from
an early episode of star formation (population III).

It is worth noting that, although the large BH mass does not require
that the emission is beamed, we cannot rule out that a moderate jet
activity, producing radio emission (and possibly inflating the
emission nebula), is present in NGC 1313 X-2 (see e.g. the case of an
ULX in NGC 5408; \citealt{kaaret03}). However, presently available
radio images of the field of NGC 1313 X-2 (Sydney University Molonglo
Sky Survey at 843 MHz and Australia Telescope Array at $\sim 5$ GHz;
\citealt{sto95}) are not sufficiently deep to allow detection.

Optical spectroscopy of object C, narrow band imaging of the nebula,
deep radio observations and an analysis of the short timescale X-ray
variability are essential to better assess the physical properties of
NGC 1313 X-2 and its environment. In particular, even a moderate
resolution spectrum of object C will make it possible to detect any
characteristic absorption and/or emission line, and then determine its
properties and redshift. These observations will allow us to
strengthen the identification of NGC 1313 X-2 with an intermediate mass
BH and foster our understanding of ULXs.

\acknowledgments

We would like to thank Valentina Braito for her valuable help with
some technical aspects involved in the analysis of {\it XMM} and {\it
ASCA} data and the Padova-Asiago Supernova Group (in particular Andrea
Pastorello) for providing us the ESO image of SN 1978K. We are also
grateful to Steve Murray for allowing the {\it Chandra} observation of
NGC 1313 X-2 to be taken as part of his GTO program, and to Albert
Kong for carefully reading the manuscript and pointing out the
possibility to reduce the {\it XMM} EPIC-PN data with the latest
release of XMM-SAS. PK acknowledges partial support from NASA grant
NAG5-7405 and Chandra grant GO2-3102X.  This work has been partially
supported also by the Italian Ministry for Education, University and
Research (MIUR) under grants COFIN-2000-MM02C71842 and
COFIN-2002-027145.


\clearpage

\begin{landscape}

\begin{deluxetable}{llllll}
\tablecolumns{7} \tablewidth{0pc} \tablecaption{Positions of NGC 1313 X-2 and
positions and optical magnitudes of field objects\label{tab0}}
\tablehead{ \colhead{Observatory/Instr.} & \colhead{Object$^a$} &
\colhead{RA[J2000]}& \colhead{DEC[J2000]}& R magnitude (Bessel-Cousins)&\colhead{Ref.}}
\startdata {\it ROSAT\/}/HRI     & NGC 1313 X-2     &  03 18 22.00$\pm$0.50
& -66 36 02.3$\pm$3.0   &       --&     \cite{sch00}\\
{\it XMM\/}/EPIC-MOS  & NGC 1313 X-2     &  03 18 22.34$\pm$0.33 &
-66 36 03.7$\pm$2.0     &       --&     \cite{mil02}\\
{\it Chandra\/}/ACIS-S& NGC 1313 X-2     &  03 18 22.18$\pm$0.12 &
-66 36 03.3$\pm$0.7     &       --&     this work   \\
ESO/3.6m      &  A    & 03 18 21.97$\pm$0.05  & -66 36 06.5$\pm$0.3 &
19.8$\pm$0.2 &
this work\\
ESO/3.6m      &  B    & 03 18 21.56$\pm$0.05  & -66 36 00.9$\pm$0.3 & 
20.7$\pm$0.2  &
this work\\
ESO/3.6m      &  C    & 03 18 22.34$\pm$0.05  & -66 36 03.7$\pm$0.3 &
21.6$\pm$0.2 &
this work\\
ESO/3.6m      &  D    & 03 18 20.96$\pm$0.05  & -66 36 03.7$\pm$0.3 &
17.8$\pm$0.2 &
this work\\
\enddata
\tablenotetext{a}{See Figure~\ref{fig3}.}
\end{deluxetable}

\begin{deluxetable}{lllcccccc}
\tabletypesize{\scriptsize}
\tablecolumns{9}
\tablewidth{0pc}
\tablecaption{Parameters of the fit of {\it ROSAT\/}, {\it ASCA\/} and
{\it XMM} observations \label{tab1}}
\tablehead{
\colhead{Observatory/Instr.} & \colhead{Obs. Id.}&
\colhead{Model} & \colhead{$N_{H}$} &
\colhead{Parameters} & \colhead{$\chi^2_{red}$
(dof)} & \colhead{$F_X^a$}&\colhead{$F^b_{0.2-10 \, {\rm keV}}$} \\
 & & & \colhead{($10^{21}$ cm$^{-2}$)} & & & \colhead{($10^{-12}$ erg cm$^{-2}$ s$^{-1}$)} & \colhead{($10^{-12}$ erg cm$^{-2}$ s$^{-1}$)} 
 }\startdata
{\it ROSAT\/}/PSPC	& rp600045n00	& Power-law  		& $1.06_{-0.50}^{+2.14}$ & $\Gamma=2.40_{-0.60}^{+1.40}$	& 0.83 (20) &  1.4  & 2.6 \\
		& 			& Blackbody  		& $0.17_{-0.15}^{+0.20}$ & $kT=0.27_{-0.03}^{+0.05}$  		& 1.01 (20) &  0.34 & 0.38\\
{\it ROSAT\/}/PSPC	& rp600504n00	& Power-law  		& $1.74_{-1.31}^{+2.95}$ & $\Gamma=2.20_{-0.90}^{+1.50}$  	& 1.43 (22) &  1.5  & 1.8 \\
		& 			& Blackbody  		& $0.09_{-0.09}^{+0.23}$ & $kT=0.35_{-0.07}^{+0.05}$  		& 1.32 (22) &  0.37 & 0.44\\
{\it ASCA\/}/GIS+SIS 	& 60028000    	& Power-law  		& $4.24_{-0.41}^{+0.41}$ & $\Gamma=1.99_{-0.09}^{+0.03}$  	& 0.99 (171)&  1.3  & 3.2 \\
             		&    		& MCD        		& $1.15_{-0.34}^{+0.35}$ & $kT=1.38_{-0.07}^{+0.07}$      	& 0.89 (171)&  2.1  & 2.6 \\
{\it ASCA\/}/GIS+SIS 	& 93010000    	& Power-law  		& $4.74_{-0.50}^{+0.50}$ & $\Gamma=2.43_{-0.07}^{+0.09}$  	& 1.28 (139)&  0.50 & 2.2 \\
             		&             	& MCD        		& $1.06_{-0.31}^{+0.31}$ & $kT=1.08_{-0.04}^{+0.04}$      	& 1.31 (139)&  0.86 & 1.0 \\

{\it XMM}/EPIC-MOS+PN 	& 0106860101    & Power-law  		& $2.69_{-0.16}^{+0.18}$ & $\Gamma=2.42_{-0.07}^{+0.07}$	& 1.20 (249)& 2.1   &	\\
			&		& MCD        		& $0.70_{-0.06}^{+0.06}$ & $kT=0.90_{-0.02}^{+0.02}$      	& 2.29 (249)& 0.86  &	\\
			&		& CompTT$^c$     		& $1.67_{-0.45}^{+0.43}$ & $kT=3.00_{-0.09}^{+0.11}$      	& 1.11 (248)& 1.2   &  	\\
      			&             	&            		&			 & $T_{0}=0.15_{-0.03}^{+0.03}$ 	&  &  &       		\\
      			&             	&            		&			 & $\tau=4.69_{-0.10}^{+0.10}$ 		&  &  &       		\\
			&		&MCD+Power-law		& $3.13_{-0.37}^{+0.92}$ & $kT=0.20_{-0.05}^{+0.04}$   		& 1.09 (247)& 2.4   &  	\\
      			&             	&            		&			 & $\Gamma=2.23_{-0.09}^{+0.15}$ 	&  &  &        		\\
			&		&MCD+Power-law$^{d}$	&$3.67_{-0.52}^{+0.94}$	 & $kT=0.20_{-0.07}^{+0.10}$   		& 1.07 (247)& 2.0   & 	\\
      			&             	&            		&			 & $\Gamma=2.23_{-0.08}^{+0.12}$ 	&  &  &        		\\
\enddata
\tablenotetext{a}{Unabsorbed flux in the 0.1--2.0 keV ({\it ROSAT}), 2.0--10.0 keV ({\it ASCA}) and 0.2--10.0 keV ({\it XMM}) energy bands}
\tablenotetext{b}{Unabsorbed flux extrapolated in the 0.2-10.0 keV band using 
the web interface to PIMMS (v 3.3)}
\tablenotetext{c}{Thermal comptonization model with Wien soft photon input}
\tablenotetext{d}{Abundance 0.5 solar}
\end{deluxetable}

\end{landscape}

\clearpage

\end{document}